\documentstyle[12pt]{article}

\title{\sc  BLACK HOLE ENTROPY }

\vspace{.5cm}
\author{{\sc valeri frolov} \vspace{.3cm} \\ {\small {\em CIAR
Cosmology
Program;  Theoretical Physics Institute,}} \\  {\small {\em
University of
Alberta, Edmonton, Canada T6G 2J1}}\thanks{E-mail address:
frolov@phys.ualberta.ca}}
\date{}

\begin{document}
\maketitle

\section{Black-Hole Entropy Problem}
According to the  thermodynamical analogy in  black hole physics,
the
entropy of a black hole  in the Einstein theory of gravity equals
$S^{BH} =A_H
/(4l_{\mbox{\scriptsize{P}}}^2)$,
where   $A_H$   is   the   area   of   a   black   hole   surface
and
$l_{\,\mbox{\scriptsize{P}}}=
(\hbar G/c^3)^{1/2}$   is   the   Planck length
\cite{Beke:72,Beke:73}.
In  black hole physics the Bekenstein-Hawking entropy $S^{BH}$ plays
essentially the same role an in the usual thermodynamics.  In
particular it
allows one to estimate what part of the internal energy of a black
hole can be
transformed into work. Four laws of black
hole physics which form the basis in the thermodynamical analogy
were
formulated in \cite{BaCaHa:73}.
The generalized second law \cite{Beke:72,Beke:73,Beke:74} (see also
\cite{ThPrMa:86,NoFr:89,Wald:92,FrPa:93} and references therein)
implies that
when a black hole is a part of the thermodynamical system the total
entropy
(i.e. the sum of the entropy of a black hole and the entropy of the
surrounding
matter) does not decrease.   The success of the thermodynamical
analogy in
black hole physics allows one to hope that this analogy may be is
even deeper
and it is possible to develop statistical-mechanical foundation of
black hole
thermodynamics.

Thermodynamical and statistical-mechanical definitions of the entropy
are
logically different.
{\it Thermodynamical entropy} $S^{TD}$ is defined by the response of
the free
energy $F$ of the system on the change of  its temperature:
\begin{equation}\label{1}
dF = -S^{TD} dT .
\end{equation}
(This definition applied to a black hole determines its
Bekenstein-Hawking
entropy.)

{\it Statistical-mechanical entropy} $S^{SM}$ is defined as
\begin{equation}\label{2}
S^{SM}=-\mbox{Tr}(\hat{\rho} \ln \hat{\rho} ) ,
\end{equation}
where $\hat{\rho}$ is the density matrix describing the internal
state of the
system under consideration.  It is also possible to introduce the
{\it
informational entropy} $S^I$ by counting different possibilities to
prepare a
system in a  final state with given macroscopical parameters from
different
initial states
\begin{equation}\label{3}
S^I =-\sum_{n}{p_n \ln p_n },
\end{equation}
with $p_n$ being the probabilities of different initial states.
In standard case all three definitions give the same answer.

Is the analogy between black holes thermodynamics and the 'standard'
thermodynamics complete? Do there exist internal degrees of freedom
of a black
hole which are responsible for its entropy?  Is it possible to apply
the
statistical-mechanical and informational definitions of the entropy
to black
holes and how are they related with the Bekenstein-Hawking entropy?
These are
the questions which are to be answered.

Historically  first attempts of the statistical-me\-cha\-nical
foundation of
the entropy of a black hole were connected with the informational
approach
\cite{Beke:73,ZuTh:85}. According to this approach the black hole
entropy is
interpreted as "the logarithm of the number of quantum mechanically
distinct
ways that the hole could have been made"\cite{ZuTh:85,ThPrMa:86}.
The so
defined informational entropy of a black hole is simply related with
the amount
of information lost by stretching the horizon, and as was shown by
Thorne and
Zurek it is equal to the Bekenstein-Hawking entropy
\cite{ZuTh:85,ThPrMa:86}.

The  dynamical origin of the entropy of a black hole and the relation
between
the statistical-mecha\-ni\-cal and Bekenstein-Hawking entropy have
remained
unclear. In the present talk I describe some new results obtained in
this
direction.

\section{Dynamical Degrees of Freedom}
The problem of the dynamical origin of the black hole entropy was
intensively
discussed recently.  The basic idea which was proposed is to relate
the
dynamical degrees of freedom of a black hole with its quantum
excitations. This
idea has different
realizations\cite{Hoof:85,FrNo:93,CaTe:93,SuUg:94,GaGiSt:94,BaFrZe:94}

\footnote{
For recent review of the problem of the dynamical origin of the
entropy of a
black hole, see  \cite{Beke:94}.
}.
Here I  discuss the recent proposal  \cite{FrNo:93,BaFrZe:94} to
identify the
dynamical degrees of freedom of a black hole with the states of all
fields
(including the gravitational one) which are propagating inside the
black hole.

In order to specify  the corresponding internal states consider at
first the
process of quantum particles creation by the gravitational field of a
black
hole. This process can be described as the effect of parametric
excitations of
zero-point fluctuations propagating in the time-dependent
gravitational field
of a  black hole.  Particles are created in pairs. The creation of a
particle
outside the event horizon is necessarily accompanied by a creation of
another
particle inside the horizon. The latter particle has negative total
energy\footnote{
It should be reminded that any isolated black hole at late time after
its
formation  is stationary. The energy is defined with respect to the
corresponding Killing vector. For a static black hole the Killing
vector is
spacelike inside the event horizon, which makes it possible the
existence of
the negative energy states.}.
 As the result of the Hawking process of pairs creation these modes
with
negative energies are permanently excited  and as we shall see later
their
state is described by a thermal density matrix. As for the particles
created
outside a black hole (external particles), only very small their
number   can
penetrate the potential barrier and reach infinity. Namely those
particles form
the  Hawking radiation of a black hole. All other external particles
are
reflected by the potential barrier and fall down into the black hole.
During
the time when they are still outside the horizon, the corresponding
internal
modes (which are described by a thermal density matrix) give the
contribution
to the black hole entropy.

In order to make the definition of the black hole entropy more
concrete
we assume that there exists a stationary black hole and denote by
$\hat{\rho}^{\mbox{\scriptsize{init}}}$ the density matrix describing
in
the Heisenberg representation  the initial state  of  quantum  fields
propagating in its background.  One may consider e.g. the in-vacuum
state for a black hole evaporating in the vacuum, or the
Hartle-Hawking
state for a black hole in equilibrium  with  thermal  radiation.
For
an exterior observer the system under  consideration consists  of two
parts: a  black hole and radiation outside of it.  The state of
radiation outside the black hole is  described   by  the   density
matrix   which  is   obtained  from
$\hat{\rho}^{\mbox{\scriptsize{init}}}$ by
averaging it over the states which are located inside the black hole
and are invisible in its exterior
\begin{equation}
\hat{\rho}^{\mbox{\scriptsize{rad}}}
=\mbox{Tr}
^{\mbox{\scriptsize{inv}}}\hat{\rho}^{\mbox{\scriptsize{init}}}.
\label{3.1}
\end{equation}
For an isolated black hole this density matrix
$\hat{\rho}^{\mbox{\scriptsize{rad}}}$
in particular describes its Hawking radiation at infinity. For a
black hole in
thermal equilibrium with radiation inside a cavity the density matrix
$\hat{\rho}^{\mbox{\scriptsize{rad}}}$ describes the state of thermal
radiation.

Analogously we define the density matrix describing the state of a
black hole as
\begin{equation}
\hat{\rho}^H =\mbox{Tr}
^{\mbox{\scriptsize{vis}}}\hat{\rho}^{\mbox{\scriptsize{init}}}.
\label{3.2}
\end{equation}
The trace-operators $\mbox{Tr} ^{\mbox{\scriptsize{vis}}}$ and
$\mbox{Tr} ^{\mbox{\scriptsize{inv}}}$ in these
relations mean that the trace  is taken over the states located
either
outside  (`visible')  or  inside   (`invisible')  the  event
horizon,
correspondingly.  We define the entropy of a black hole as
\begin{equation}
S^H =-\mbox{Tr} ^{\mbox{\scriptsize{inv}}}(\hat{\rho}^H \ln
\hat{\rho}^H ).
\label{3.3}
\end{equation}

The proposed definition of  the entropy of  a black hole  is similar
to the
definition of the entropy of a usual black body. The definition is
invariant
in   the  following   sense.    Independent   changes   of
vacuum
definitions  for `visible' and  `invisible' states  do not  change
the
value   of   $S^H$.    Bogolubov's   transformations    describing
an
independent changes  of the vacuum states inside and outside  the
black
hole  can  be   represented  by  the         unitary
operator
$\hat{U}=\hat{U}^{\mbox{\scriptsize{vis}}}
\otimes \hat{U}^{\mbox{\scriptsize{inv}}}$,        where
$\hat{U}^{\mbox{\scriptsize{vis}}}$         and
$\hat{U}^{\mbox{\scriptsize{inv}}}$ are   unitary
operators   in  the  Hilbert  spaces    of `visible'   and
`invisible'
particles,   correspondingly.    The above  used trace  operators
are
invariant under such transformations.

In order to define the states one usually use modes expansion.  The
modes are
characterized by a complete set of quantum numbers. Due to the
symmetry
properties  one can choose such a subset $J$ of quantum numbers
connected with conservation laws (such as orbital and azimuthal
angular momenta, helicity and so on) that
guarantees the factorization of the density matrices.
In the absence of mutual interaction of
different fields the subset $J$ necessarily includes  also the
parameters
identifying the type of the field (e.g. mass, spin, and charge).
The factorization in particular means that
\begin{equation}
\hat{\rho}^{\mbox{\scriptsize{init}}}=\otimes _J
\hat{\rho}^{\mbox{\scriptsize{init}}}_J ,\label{3.4}
\end{equation}
where $\hat{\rho}^{\mbox{\scriptsize{init}}}_J$
is acting in the Hilbert space ${\cal H}_J$ of states with the chosen
quantum numbers $J$, while the complete Hilbert space is ${\cal
H}=\otimes _J {\cal H}_J$. The factorization also means that the
separation into `visible' and `invisible' states can be done
independently in each subset of modes with a fixed $J$ so that
\begin{equation}
\hat{\rho}^H =\otimes _J \hat{\rho}^H_J \hspace{1cm} \hat{\rho}^H_J
=\mbox{Tr}^{\mbox{\scriptsize{vis}}}_J
\hat{\rho}^{\mbox{\scriptsize{init}}}_J,
\label{3.5}
\end{equation}
where all the operators with subscript $J$ are acting in the Hilbert
space
${\cal H}_J$\footnote{
Instead of non-normalizable monochromatic waves $f_{\omega lm}$ with
fixed
frequency $\omega$ it is convenient to consider normalized wave
packets
constructed from the monochromatic waves
\[
f_{jnlm}=\delta ^{-1/2}\kappa^{-1}\int ^{(j+1)\delta}_{j\delta}
\exp (2\pi in\kappa^{-1}\omega /\delta )f_{\omega lm}d\omega ,
\]
where  $0<\delta\ll 1$. In what follows we shall be working with
these type
wavepackets and assume that the collective index  $J=(j,n,l,m)$. We
denote by
$\lambda$ the collective index $\lambda=(\omega,l,m)$.
}.

We illustrate the main steps of the calculations of the contribution
of a given
field to the density matrix of a black hole for a
spherically-symmetric black
hole. (The presence of charge and rotation does not create problems.)
Moreover for simplicity we assume that a black hole is surrounded by
a
spherical mirror-like boundary of size $r_B<3M$,  so that the black
hole is in
thermal equilibrium with the radiation inside the cavity and the
state of the
system is described by the Hartle-Hawking vacuum state.  The easiest
way to
introduce this state as well as to give definition of the 'up'- and
'side'-modes we use to describe the states of  'visible' and
'invisible'
particles one can use the following useful trick proposed by Hawking
in his
original paper on the black hole evaporation \cite{Hawk:75}. At late
time after
the formation of a black hole the spacetime   with the high accuracy
is
described by a stationary (in our case static) metric.
For a given black hole we define its 'eternal version' as a spacetime
of an
eternal black hole  which has the same global parameters (mass,
charge, angular
momentum) as the given black hole. Modes of the field $\phi$
propagating at
late-time in an 'original' black hole can be traced back in time in
the
spacetime of its 'eternal version' up to the initial global Cauchy
surface
$\Sigma$ described by the equation $t=0$, where $t$ is the global
time
parameter defined by the Killing vector. 'Up'-modes are defined as
positive (in
time $t$) solutions vanishing in the left wedge $R_-$, and
'side'-modes are
defined as negative (in time $t$) solutions vanishing in the right
wedge $R_+$.
In the presence of mirror-like boundaries $B$ and $B'$ they are to
satisfy the
corresponding boundary conditions at $B$ and $B'$\footnote{
For more details concerning the definition of 'up' and 'side' modes
and their
relations with other standard modes ('in', 'out', and 'down') defined
in a
black hole's exterior, see \cite{BaFrZe:94}
}.

By using the linear combinations of the operators of creation and
annihilation
for 'up'- and 'side'-particles  one can construct the operators which
annihilate the Hartle-Hawking vacuum $|HH\rangle$.   For our choice
of the
initial state $\hat{\rho}^{\mbox{\scriptsize{init}}}
=|HH\rangle\langle HH|$.
This density matrix can be evidently expressed as the function of the
'up' and
'side' creation and annihilation operators.  In order to separate the
states
into 'visible' and 'invisible' we consider a spacelike or null
surface
$\Sigma_0$ which cross the event horizon at late time after black
hole
formation. An 'up'-particle is 'visible' if it crosses  the surface
$\Sigma_0$
of a chosen moment of time outside the horizon. For this mode $J$,
the density
matrix $\hat{\rho}^H_J$, describing the state of the corresponding
'side'-particle, is obtained by tracing
$\hat{\rho}^{\mbox{\scriptsize{init}}}_J$ over the Hilbert space of
'up'-particles with index $J$. As the result one obtains the thermal
density
matrix of the form \cite{FrNo:93}
\begin{equation}
\hat{\rho}^H_J =\rho ^0_J \exp \left[-(\omega _J/T_H )
\hat{\alpha}_J^{*SIDE} \hat{\alpha}_J^{SIDE}\right].   \label{3.6}
\end{equation}
In other words Hawking radiation of a black hole is accompanied by
thermal
excitation of 'side' modes (black hole's internal degrees of
freedom).
For the 'up'-modes $J$ crossing  $\Sigma_0$  inside the horizon, both
'up' and
'side' particles  are 'invisible'. We denote by
$\hat{\rho}^{\bullet}_J$, the
density matrix for such a pair . The density matrix
$\hat{\rho}^{\bullet}_J$
evidently describes the pure state, so that such a pair does not
contribute to
the entropy of a black hole.  For the total density matrix
$\hat{\rho}^H$ of a
black hole we have the following representation
\begin{equation}\label{3.7}
\hat{\rho}^H=\prod_{J}{}^{'}{\hat{\rho}^H_J}\prod_{J}
{}^{''}{\hat{\rho}^{\bullet}_J} .
\end{equation}
Here the prime indicates that the product includes only those states
for which
the 'up'-modes are 'visible', and double prime indicates that the
product
includes  states for which the 'up'-modes are 'invisible'.

The expression (\ref{3.6}) was obtained for the special choice of the
initial
state.  For another choice of the initial state this expression  must
be
modified. What is important that the modifications practically do not
influence
 the expression for $\hat{\rho}^H_J$  for late-time modes $J$.  A
'side'-mode
propagating inside a black hole close to the horizon at late-time
being traced
back in time reaches $\cal J^-$ with a huge blue shift proportional
to $\exp
\kappa t$, where $t$ is time past after the formation of the black
hole. It
means that in order to change the distribution for the mode $J$ (e.g.
to add an
additional quantum in this state) one needs to send from $\cal J^-$
the
excitation which has exponentially large energy. At sufficiently late
time
this energy for given frequency is much larger than the mass of the
black hole.
For this reason  the density matrix for 'side'-modes at late time
will have the
universal form (\ref{3.6}).

\section{Statistical-Mechanical Entropy}
By using the density matrix of a black hole $\hat{\rho}^H$ one can
calculate
the corresponding statistical-mechanical entropy
$S^{SM}=-\mbox{Tr}(\hat{\rho}^H\ln \hat{\rho}^H)$. The main
contribution to the
entropy of a black hole is given by 'side' modes of fields located
in the very
close vicinity of the horizon. Contributions of different fields
enter $S^{SM}$
additively. The factorization property (\ref{3.5}) allows to write
\begin{equation}
S^{SM} =\sum_J
S_J^{SM} ,\hspace{1cm} S_J^{SM} =-\mbox{Tr}
^{\mbox{\scriptsize{inv}}}_J
(\hat{\rho}^H_J \ln
\hat{\rho}^H_J ), \label{4.1}
\end{equation}
where all the operators with subscript $J$ are acting in the Hilbert
space
${\cal H}_J$. In accordance with this decomposition one has\footnote{
We use units in which $G=c=\hbar=k=1$.
}
\begin{equation}\label{4.2}
S^{SM}={\sum_{J}}^{'}{s(\beta \omega_J )},
\end{equation}
where
\begin{equation}\label{4.3}
s (\beta\omega ) = {{\beta\omega } \over {e^{\beta\omega }-1}} -
	\ln (1- e^{-\beta\omega }),
\end{equation}
is  the entropy of a single oscillator with the frequency $\omega$ at
temperature $T=1/\beta$
and the prime in sum indicates that the summation is taken over the
modes for
which the corresponding 'up'-particle is 'visible' at the chosen
moment of time
$\Sigma_0$. The terms $\hat{\rho}^{\bullet}_J$ corresponding to a
pure state of
a pair inside a black hole do not contribute to $S^{SM}$.

Returning from wave packets $J$ to monochromatic waves $\lambda$ one
obtains
 the following result for the contribution of a  chosen field to
$S^{SM}$
\begin{eqnarray} \label{4.4}
 S^{SM} = \int {d {\mbox{\boldmath $x$}}}
        \sum_{\lambda}  \mu _\lambda  ({\mbox{\boldmath $x$}})
s(\beta\omega_{\lambda}) .
	\end{eqnarray}
Here
	\begin{eqnarray}
	\mu _\lambda  ({\mbox{\boldmath $x$}}) = g^{\tau\tau} g
^{1/2}
	[R_\lambda ({\mbox{\boldmath $x$}})]^2
	\end{eqnarray}
is a  phase space density of quantum modes and  $R_\lambda
({\mbox{\boldmath
$x$}})$ are spatial harmonics corresponding to the mode with a
collective
quantum number $\lambda=(\omega,l,m)$.

The so defined $S^{SM}$ contains a divergence, connected with the
integration
over the space regions near the horizon and is of the form
\cite{FrNo:93}
\begin{eqnarray}\label{4.6}
S^{SM}\approx \frac{\alpha}{\varepsilon}
\end{eqnarray}
where $\alpha$ is dimensionless parameter depending on the type of a
field,
$\varepsilon =(l/r_+)^2$,  and $l$ is the proper distance cut-off
parameter.
For a conformal scalar massless field $\alpha =1/90$.  For a fixed
value $l$ of
the cut-off parameter, the expression for $S^{SM}$ does not depend on
the
particular choice of the surface $\Sigma_0$, which was introduced to
define
'visible' and 'invisible' particles. One may expect that  quantum
fluctuations
of the horizon may provide natural cut-off and make $S^{SM}$ finite.
Simple
estimations of the cut-off parameter show that $l\approx
l_{\,\mbox{\scriptsize{P}}}$ and $S^{SM} \approx S^{BH}$
\cite{FrNo:93}\footnote{
The loop expansion  can be formulated as the expansion in powers of
$\hbar$,
and hence $S^{SM}$, which is one-loop quantity, must contain extra
$\hbar$ with
respect to the tree-level contribution $S^{BH}$. Nevertheless the
calculations
show that $S^{SM}$  and $S^{BH}$  are of the same order of magnitude.
It
happens because the cut-off parameter needed to make $S^{SM}$ finite
is also
dependent on $\hbar$: $\varepsilon\sim \hbar$.
}.

\section{No-Boundary Wave Function}
Another approach to the problem of dynamical degrees of freedom of a
black hole
was proposed in Ref.\cite{BaFrZe:94}. Its basic idea  is the
following. The
study  of propagation of perturbations of physical fields in the
spacetime of a
real black hole can be
reduced to the analogous problem for its 'eternal version' .
In particular one can trace back in time the perturbations  in the
space of the
'eternal version' until they cross the section $\Sigma$ described by
the
equation $t=0$. This section is known as the Einstein-Rosen bridge.
The number
and properties of the physical fields depends on the particular
model. In any
case the gravitational field must be included. The initial data at
$\Sigma$ for
the gravitational perturbations at $\Sigma$ can be related with small
deformations of the geometry of the Einstein-Rosen bridge.
The space of  physical configurations
of a system including a black hole can be related to the space of
'deformations' of the Einstein-Rosen bridge of the eternal black hole
and
possible configurations of other (besides the gravitational) fields
on it,
which
obey the constrains and preserve asymptotic flatness.   In a
spacetime of an
'eternal
version' of a black hole,  perturbations with initial data located on
the inner
part of the Einstein-Rosen bridge ($\Sigma_-$) are propagating to the
future
remaining
entirely inside the horizon, and hence the corresponding
perturbations in a 'physical'  black hole also always  remain under
the
horizon. That is why these data should be identified
with internal degrees of freedom of a black hole.  This construction
allows
natural generalization to the case when the deformations of the
Einstein-Rosen
bridge are not small. A quantum state of a black hole can be
described by a
wavefunction   defined as a functional on the configuration space of
deformations of the Einstein-Rosen bridge. In this representation
deformations
of the external ($\Sigma_+$) and internal ($\Sigma_-$) parts of the
Einstein-Rosen bridge naturally represent degrees of freedom of
matter outside
the black hole and black hole's internal degrees of freedom.

The no-boundary ansatz (analogous to Hartle-Hawking ansatz in quantum
cosmology) singles out a state which plays the role of a ground state
of the
system \cite{BaFrZe:94}.  In order to describe this ansatz for a
black hole we
consider a half of the Hawking-Gibbons instanton , i.e. the  space of
the
Euclidean black hole with the metric (\ref{2.0b}) and $\tau\in
(0,\pi/\kappa)$
Besides the boundary at infinity $\partial\mbox{\boldmath
$M$}_{\infty}$ this
Euclidean space $\mbox{\boldmath $M$}$ possesses only one boundary
$\partial\mbox{\boldmath $M$}$ which is isometric to the
Einstein-Rosen bridge.

The no-boundary wavefunction of a black hole \cite{BaFrZe:94} is
defined  as a
path integral
	\begin{eqnarray}
	\mbox{\boldmath $\Psi$}(^3\!g(\mbox{\boldmath $x$}),
	\varphi(\mbox{\boldmath $x$}))=
	\int {\cal D} \,{}^4\!g \ {\cal D}\mbox{\boldmath $\phi$}\
	{\rm e}^{\!\!\phantom{0}^
	{\textstyle -\mbox{\boldmath $I$}
	[\,{}^4\!g ,\mbox{\boldmath $\phi$}\,]}}
\label{5.1}
	\end{eqnarray}
of the exponentiated gravitational action $\mbox{\boldmath
$I$}[\,{}^4\!g
,\mbox{\boldmath $\phi$}\,]$ over Euclidean 4-geometries and
matter-field
configurations on  those spacetime histories of physical  fields
$\phi=\phi(x)$ on $\mbox{\boldmath $M$}$ that generate the Euclidean
4-geometries asymptotically flat at
the infinity $\partial\mbox{\boldmath $M$}_{\infty}$ of spacetime
and  are
subject to the
conditions
	$(^3\!g(\mbox{\boldmath $x$}),\,
	\varphi(\mbox{\boldmath $x$})),\;\;
	\mbox{\boldmath $x$}
	\in\partial\mbox{\boldmath $M$}$, -- the collection of
3-geometry and boundary
matter fields  on $\partial\mbox{\boldmath $M$}$, which are just the
argument
of the wavefunction (\ref{5.1}).  $\mbox{\boldmath
$I$}[\,{}^4\!g ,\mbox{\boldmath $\phi$}\,]$ is the Lagrangian
gravitational
action in terms of these
fields. The integration measure ${\cal D} \,{}^4\!g \ {\cal D}\phi$
involves
the local functional
measure the structure of which is not very important for our
purposes.

 By its construction the no-boundary wavefunction
 of a black hole is symmetric with respect to the transposition of
the interior and exterior parts of the Einstein-Rosen bridge.  We
call this
property {\em duality}.   For a 'real' black hole  formed in the
gravitational
collapse, this exact symmetry is broken. Nevertheless, since there
is a
close relation between physics of a 'real' black hole and its
'eternal
version',  the duality of the above type plays an important role and
allows
one, for example, to explain why the approach based on identifying
the
dynamical degrees of freedom of a black hole with its external modes
gives
formally the same answer for the dynamical entropy of a black hole as
our
approach.

In the semiclassical approximation the no-boundary wavefunction
Eq.(\ref{5.1})
takes the form
	\begin{eqnarray}
	\Psi(\varphi,M)=P\,{\rm e}^{\!\!\phantom{0}^{
	{\textstyle - 2\pi M^2\! - \mbox{\boldmath $I$}_2[\,\phi
(\varphi)\,]}}},
     \label{5.2}
	\end{eqnarray}
where $\mbox{\boldmath $I$}_2[\,\phi (\varphi)\,]$ is a quadratic
term of the
action in the linearized physical fields, and  $\phi (\varphi)$ is a
solution
of the corresponding field equations on $\mbox{\boldmath $M$}$
matching the
boundary conditions $\varphi$ on  $\partial\mbox{\boldmath $M$}$.
Denote by  $\varphi_{\lambda,\pm}$  the coefficients in
 the decomposition  of the field $\varphi ({\mbox{\boldmath $x$}})$
on
$\Sigma_{\pm}$ (an external and internal parts of the Einstein-Rosen
bridge
$\Sigma=\partial\mbox{\boldmath $M$}$)
in the basis of spatial harmonics
$\varphi_{\pm}({\mbox{\boldmath $x$}})=
	\sum_{\lambda} \varphi_{\lambda,\pm}
R_{\lambda}({\mbox{\boldmath $x$}}).$
Then the quadratic form $\mbox{\boldmath $I$}_2[\,\phi (\varphi)\,]$
is
	\begin{eqnarray}
\hspace{-.3cm}\mbox{\boldmath $I$}_2(\varphi_{+},\varphi_{-})=
	\frac12\,\sum_{\lambda}\left
	\{\,\frac{\omega_{\lambda}\,{\rm
cosh}(\beta_0\,\omega_{\lambda}/2)}
	{{\rm sinh}(\beta_0\,\omega_{\lambda}/2)}
	\,(\varphi_{\lambda,+}^2+\varphi_{\lambda,-}^2)
	-\frac{2\,\omega_{\lambda}}{{\rm sinh}(\beta_0\,
	\omega_{\lambda}/2)}
	\,\varphi_{\lambda,+}\varphi_{\lambda,-}\!\right\}.
\label{5.3}
	\end{eqnarray}

The dependence of the no-boundary wavefunction on the mass $M$ shows
in this
state  the it is most probable to find a black hole of the smallest
possible
(Planckian) mass. For other states
the dependence of the probability on the mass parameter is different.
One might
expect
that for a state with fixed average value $M_0$ of energy (mass)  the
corresponding
mass dependent part of the probability will have the form $\exp [-
(M-M_0)^2/2\Delta^2]$, where $\Delta$ is the mass dispersion.

For study the fields contribution to the statistical-mechanical
entropy in the
one-loop approximation it is sufficient to fix mass $M$ of a black
hole as a
parameter in the wave function, and consider only the  part
describing fields
perturbations. It is possible to show that for a fixed mass parameter
$M$ the
part $\exp{ - \mbox{\boldmath $I$}_2(\varphi_{+},\varphi_{-})}$ of
the
no-boundary wave function (\ref{5.2}) with $\mbox{\boldmath
$I$}_2(\varphi_{+},\varphi_{-})$ given by (\ref{5.3}) describes the
Hartle-Hawking vacuum state of the corresponding perturbations
$\phi$.

By tracing over the external variables one obtains the density matrix
of a
black hole  and can calculate its statistical-mechanical entropy
\cite{BaFrZe:94}. The result  coincides with (\ref{4.6}). The reason
why the
calculations based on a 'real' black hole and on its 'eternal
version' give the
same result for the statistical-mechanical entropy of a black hole is
the
following. As it was already mentioned the late-time occupation of
'side'-modes
does not depend of the particular choice of the initial state, so
that  to
compare the results of the calculations one can choose the
Hartle-Hawking state
for the 'real' black hole, which is virtually the same as for the
no-boundary
wave function.  On the other hand the main contribution to the
entropy is
connected with modes, propagating in the very close vicinity of the
horizon,
which are highly blue-shifted and for which the geometric-optics
approximation
works extremely well when one propagates these modes to the
Einstein-Rosen
bridge $\Sigma$ in the eternal version of a black hole. For these
reasons the
counting of modes, contributing to the statistical-mechanical entropy
for a
'real' black hole and for its 'eternal version' give the same answer.

\section{Thermodynamical Entropy of a Black Hole}
The Bekenstein-Hawking entropy  by its definition coincides with the
thermodynamical entropy of a black hole. We discuss now the relation
between
the thermodynamical and statistical-mechanical entropies and show
that for a
black hole these entropies are different.

In order to derive thermodynamical characteristics of a black hole it
is
convenient to
begin with the partition function $Z(\beta)$. It is related with the
free
energy $F$ ($Z(\beta)=\exp(-\beta F)$) and is defined by the
functional
integral\cite{Hawk:79,BrYo:93}
\begin{equation}\label{8}
Z(\beta )=\int D[g,\phi]\exp (i I [g,\phi ]) ,
\end{equation}
where $I [g,\phi ]$ is the action for the gravitational field $g$ and
some
other fields $\phi$.
The state of the system is determined by the choice of the boundary
conditions
on the metrics and fields that one integrates over. For the canonical
ensemble
describing the gravitational fields within a spherical box of radius
$r_B$ at
temperature $T_B$ one must integrate over all the metrics inside
$r_B$ which
are periodically identified in the imaginary time direction with
period
$\beta_B=T_B^{-1}$. Denote by $(g_0,\phi_0)$ a point of the extremum
of the
action $I [g,\phi ]$, then
\begin{equation}\label{9}
\ln Z =iI[g_0,\phi_0]+\ln \int D[\bar{g},\bar{\phi}]\exp (i I_2
[\bar{g},\bar{\phi }])
\end{equation}
where $\bar{g}_{\mu\nu}=g_{\mu\nu}-g_{0\mu\nu}$, $\bar{\phi}=\phi
-\phi_0 $,
and $I_2 [\bar{g},\bar{\phi }]=I [g,\phi ]-I[g_0,\phi_0]$ is
quadratic in the
perturbations $\bar{g}$ and $\bar{\phi}$.

For vanishing background field $\phi_0 =0$ the extremum $g_0$ is a
solution of
the vacuum Einstein equations. This solution for given boundary
conditions
coincides with the Euclidean black hole (a Hawking-Gibbons
instanton). The
corresponding metric can be obtained from the Schwarzschild solution
\begin{eqnarray}\label{2.0a}
ds^2 &=&-Bdt^2 +B^{-1}dr^2 +r^2 d\omega^2,  \\
d\omega^2&=&d\theta^2+\sin^2\theta d\varphi^2 ,\hspace{.6cm} B=1-2M/r
,
\end{eqnarray}
by the  Wick's rotation of time $t\rightarrow -i\tau$. The metric of
the
Euclidean black hole is
\begin{eqnarray}\label{2.0b}
ds^2 =Bd\tau^2 +B^{-1}dr^2 +r^2 d\omega^2 .
\end{eqnarray}
This metric is regular at the horizon $r=r_+=2M$ (where $B=0$) only
provided
$\tau$ is periodic with the period $2\pi/\kappa$ ($\kappa$ is the
surface
gravity of a black hole, for the Schwarzschild black hole
$\kappa=1/4M$). The
property of periodicity with respect to the imaginary time $it$ with
the period
$\beta=2\pi/\kappa$  implies in particular that a black hole is in
thermal
equilibrium with surrounding thermal radiation of the field $\phi$,
provided
the temperature of radiation, measured at infinity, is
$T_{BH}=\beta^{-1}=\kappa/2\pi$.

For this extremum  $iI[g_0,\phi_0]=-I_E[g_0]$, where $I_E$ is the
Euclidean
action. The relation (\ref{9}) implies that the free-energy $F$ can
be written
as
\begin{equation}\label{10}
F=F_0+F_1+\ldots ,
\end{equation}
where $F_0 =\beta^{-1}I_E[g_0]$ and $F_1$ are the tree-level and
one-loop
contributions, respectively, and dots denote higher order terms  in
loops
expansion.

We demonstrate now that the tree-level part of the free energy is
directly
connected with the Bekenstein-Hawking entropy, while the
statistical-mechanical
entropy is related to the one-loop contribution $F_1$. In what
follows we
assume that a black hole is in equilibrium with the thermal
radiation inside the cavity of radius $r_B$. In the thermal
equilibrium the
mass $M$  of a black hole is related to $\beta$ as $\beta=4\pi r_+$,
where
$r_+ =2M$ is the gravitational radius.The equilibrium is stable if
$r_B <3r_+
/2$. The tree-level contribution of the black hole to the free energy
of the
system  can be written in the form \cite{York:86,BrBrWhYo:90}
\begin{equation}\label{11}
F_0 = r_B \left( 1-\sqrt{1-r_+ /r_B}\right) -\pi
r_+^2\beta_B^{-1} ,
\end{equation}
where $\beta_B=4\pi r_+
(1-r_+ /r_B )^{1/2}$ is the inverse temperature at the boundary
$r_B$. The
tree-level contribution $S_0^{TD}$ to the thermodynamical entropy of
a black
hole
defined as $S_0^{TD} =-\partial_{T_B}F_0\equiv
\beta_B^2\partial_{\beta_B}F_0$ is $S_0^{TD} =\pi r_+^2\equiv
A/4l_P^2\equiv S^{BH}$. In addition to this tree-level contribution
which
identically coincides with the Bekenstein-Hawking entropy $S^{BH}$
there are
also
one-loop contributions directly connected with dynamical degrees of
freedom of
the black hole, describing its quantum excitations. We consider them
now in
more details.

By using Eq.(\ref{9}) the one-loop contribution $F_1$ to the free
energy can be
written in the form $F_1=-\beta^{-1}\ln Z_1$, where
\begin{equation}\label{12}
Z_1=\int D[\varphi] \exp (-I_E[\varphi]),
\end{equation}
and $I_E[\varphi]$ is the quadratic Euclidean action of the field
configuration
$\varphi\equiv (\bar{g},\bar{\phi})$. The integration is performed
over all the
perturbations fields $\varphi$ that are real on the Euclidean section
with
metric (\ref{2.0b}) and are periodic in imaginary time coordinate
$\tau$ with
period $\beta$\footnote{ For a black hole  located inside a cavity
of radius
$r_B$ the field $\varphi$
must obey also some boundary conditions at $r_B$.  For our
consideration
dependence of $F_1$ on $r_B$ is not important. That is why we do not
indicate
it explicitly. We also use the inverse temperature at infinity
$\beta$ instead
of $\beta_B$.
} . In the one-loop approximation different fields give independent
contributions to $F_1$. For this reason it is sufficient to calculate
the
contribution of a chosen field $\varphi$ and then add all the
contributions
corresponding to different fields. The integral (\ref{12}) is
ultraviolet
divergent and requires regularization.  The regularized value of
$F_1$ may
depend on some regularization mass parameter
$\mu$\cite{Alle:86,DoKe:78}.
Below we assume that the corresponding regularization is made and use
the
notation $Z_1$ and $F_1$ for the renormalized values of these
quantities.

According to its definition the one-loop contribution $S_1^{TD}$ to
the
thermodynamical entropy of a black hole is determined by the total
response of
the one-loop free energy $F_1$ on the change of the temperature.
Besides the
direct dependence of $F_1$ on temperature it also depends on the mass
$M$ of a
black hole. In the thermal equilibrium $M$ is a function of
temperature. Thus
we have ($r_+=2M$)
\begin{equation}\label{13}
S^{TD}_1=\beta^2{d{F_1} \over d\beta}\equiv \beta^2 \left.{\partial
{F_1}\over
\partial \beta}\right| _{r_+}+\beta^2\left.{\partial {F_1}\over
\partial
r_+}\right|_{\beta}{d{r_+} \over d\beta} .
\end{equation}
The first term in the right-hand-side of this relation is equal to
the one-loop
contribution $S_1^{SM}$ to the statistical-mechanical entropy. In
order to
justify this claim we use the fact that  the partition function
$Z_1$ is
related to the thermodynamical partition function $Z^T(\beta)$ of the
canonical
ensemble
\begin{equation}\label{14}
Z^T(\beta)={\mbox {Tr e}}^{-\beta \hat{H}}=\sum \exp (-\beta E_n ) ,
\end{equation}
where $E_n$ is the energy (eigenvalue of the Hamiltonian $\hat{H}$ of
the field
$\varphi$). Namely,  Allen\cite{Alle:86}  showed that  $F_1\equiv
-\beta^{-1}\ln Z_1$  differs from $F^T\equiv -\beta^{-1}\ln Z^T$ only
by terms
which are  independent of  $\beta$\footnote{
Recently the paper by D.V.Fursaev appeared as the preprint DSF-32/94
(hep-th/9408066), in which the problem of renormalization on the
manifolds with
cone-like singularity was considered. In this paper it was argued
that in the
presence of cone singularities additional temperature dependent
divergencies
may arise. If it happens the relation between $Z_1$ and $Z^T$
obtained by
Allen\cite{Alle:86} might be modified. As the result of this
modification
$\Delta S_1$ in Eq.(\ref{15}) would get extra contribution. However
the main
conclusions of the present paper remains unchanged. The reason is
that for
$\beta=\beta_H$ the cone singularity disappears, the extra surface
terms in the
effective action vanish, and the renormalized free energy  remains
finite. So
that the mechanism of compensation discussed later in the present
paper remains
valid.
}.  Hence we have $\beta^2 ({\partial {F_1}/ \partial
\beta})_{r_+}=\beta^2
({\partial {F^T}/ \partial
\beta})_{r_+}=-\mbox{Tr}(\hat{\rho}\ln{\hat{\rho}})\equiv S_1^{SM}$,
where
$\hat{\rho}=\exp[-\beta (\hat{H}-F^T)]$. (We used here that the
partial
derivative $({\partial / \partial \beta})_{r_+}$ with respect to the
inverse
temperature $\beta$  commutes with $\mbox{Tr}$-operation.)
The above relations allow one to rewrite Eq.(\ref{13}) in the form
\begin{equation}\label{15}
S^{TD}_1=S^{SM}_1+\Delta S_1 ,
\end{equation}
where $\Delta S_1\equiv \beta^2 ({\partial {F_1}/ \partial {
r_+}})_{\beta}
dr_+/d\beta$. This relation shows that in order to obtain $S^{TD}_1$
the
statistical-mechanical entropy must be 'renormalized' by adding
$\Delta S_1$.
In particular, the relation (\ref{15}) may give an explanation to the
entropy
renormalization procedure proposed by Thorne and Zurek\cite{ZuTh:85}.

For an investigation of $\Delta S_1$,  it is convenient to  rewrite
$F_1$ which
enters the definition of  $\Delta S_1$ as $F_1=(F_1-F^T)+F^T$. The
difference
$F_1-F^T$ does not depend on $\beta$ and hence one can calculate its
value for
zero temperature ($\beta=\infty$). It indicates that the
corresponding
contribution to $\Delta S_1$ is directly connected with vacuum
polarization.
The second contribution to $\Delta S_1$ (connected with $F^T$ term)
arises
because  the complete derivative $d/d\beta$, defined by the relation
(\ref{13}), does not commute with $\mbox{Tr}$-operation\footnote{
Possible non-commutation of differentiaton with respect to
temperature and
Tr-operation for thermodynamical systems in the framework of
thermofield
approach was discussed in Ref.\cite{EvHaUmYa:93} }. It is instructive
to
demonstrate in more detail the origin of this non-commutation. The
thermodynamical partition function $Z^T$ in a static spacetime can be
presented
in the form\cite{Alle:86}
\begin{equation}\label{16}
Z^T=\prod_{\lambda}{\left[1-\exp(-\beta\omega_{\lambda})\right]
^{-1}},
\end{equation}
where $\omega_{\lambda}$ are the energies of the single-particle
states (or
modes) and $\lambda$ is the index enumerating these states. For the
free energy
$F^T$ we have
\begin{equation}\label{17}
F^T =\sum_{\lambda}{f(\beta\omega_{\lambda}})=\int{d\omega
N(\omega|r_+)}f(\beta\omega) ,
\end{equation}
where $f(\beta\omega)=\beta^{-1}\ln [1-\exp (-\beta\omega )]$, and
$N(\omega
|r_+)$ is the density of number of states at the given energy
$\omega$ in a
black hole of mass $M=r_+/2$.  $dN/d\beta\ne 0$, since $N(\omega
|r_+)$ depends
on the mass of a black hole\footnote{
It is easy to show that $N(\omega |r_+)=\int d{\mbox{\bf
x}}\sum_{\lambda}\delta(\omega-\omega_\lambda)|g^{tt}g^{1/2}
|R_{\lambda}({\mbox{\bf x}})|^2$, where $R_{\lambda}({\mbox{\bf x}})$
are spatial harmonics\cite{BaFrZe:94}. The spatial integral is
divergent near the

horizon and requires cut-off. The main (leading at the horizon) part

of $N(\omega |r_+)$ can be calculated exactly. For example, for a

scalar massless field $N(\omega |r_+)\sim 8\omega^2 r_+^3/(\pi
\varepsilon)$,

where $\varepsilon=(l/r_+)^2$ is a dimensionless cut-off parameter
($l$

is a proper-distance cut-off).
}. This implies that $d/d\beta$ and $\mbox{Tr}$-operation do not
commute.

\section{Why the Entropy is $A/4$?}
The  calculation  of the quantities which enter Eq.(\ref{15}) is
quite
complicated. But  important conclusions can be easily obtained by
using some
general properties of the free energy $F_1$.
In general case (if $\beta\ne\beta_{H}=4\pi r_+$) the free energy
$F_1$
contains a  divergence connected with the space integration over the
region
near the horizon.  In order to regularize this divergence we suppose
that the
integration is performed up to the proper distance $l$ to the
horizon. Denote
$\varepsilon=(l/r_+)^2$. In order to emphasize the dependence of
$F_1$ on the
dimensionless cut-off parameter $\varepsilon$ we shall write
$F_1=F_1(\beta
,r_+ ,\varepsilon )$. The free energy has the same dimension as $
r_+^{-1}$ and
hence it can be
presented in the form
$F_1(\beta ,r_+ ,\varepsilon )\equiv r_+^{-1} {\cal F} (\beta/\beta_H
,\varepsilon )$,
where ${\cal F}$ is dimensionless function of two dimensionless
variables and
$\beta_H=4\pi r_+=1/T_H$,  ($T_H$ is the black hole temperature).
The
structure of the divergence near the Euclidean horizon can be
analysed by using
the curvature expansion of  $F_1$. The leading
divergent near the horizon $r=r_+$ term is
\begin{equation}\label{18}
{\cal F} (\beta/\beta_H,\varepsilon )\approx
-\varepsilon^{-1}f(\beta/\beta_H).
\end{equation}
An explicit form of the function $f$ can be obtained by analyzing the
free-energy in a flat cone space. The high-temperature
expansion\footnote{
Dowker and Kennedy have shown  that in the framework of this
expansion the
temperature $T$ naturally enters in the combination
$T/\sqrt{|g_{tt}|}$. For
this reason the expansion can be used to get more detailed
information about
the behavior of the free-energy near the horizon.
}
 (see, e.g. paper by Dowker and Kennedy\cite{DoKe:78}) shows that
$f(x)\sim
x^{-4}$ for $x\rightarrow\infty$.
The divergence of  $F_1$ at the Euclidean horizon  for
$\beta\ne\beta_H$
reflects the fact that the
number of modes that contribute to the free energy and entropy is
infinitely
growing as one considers regions closer and closer to the horizon
\cite{FrNo:93}.
For $\beta=\beta_H$ the metric (\ref{2.0b}) is regular at the
Euclidean horizon
and hence the renormalized free energy calculated for the regular
Euclidean
manifold is finite. It implies that $f(1)=0$.

The one-loop contribution of a quantum field $\varphi$ to the
statistical-mechanical entropy is
\begin{equation}\label{21}
S_1^{SM} =\left[ \beta^2 {{\partial {F_1}}\over {\partial \beta
}}\right] _{\beta =\beta_H} .
\end{equation}
It should be stressed that one must put $\beta=\beta_H$ only after
the
differentiation. The leading (divergent near the horizon) term of
$S_1^{SM}$ is
\begin{eqnarray}\label{22}
S_1^{SM}\approx -\frac{4\pi}{\varepsilon} f'(1).
\end{eqnarray}
This relation reproduces Eq.(\ref{4.6}) with $\alpha=-4\pi f'(1)$.
For a conformal massless scalar field $f'(1)=-1/(360\pi)$. If the
proper-distance cut-off parameter $l$ is of the order of  the Planck
length
$l_P$ then the contribution of the field to the
statistical-mechanical entropy
of a black hole is of the order $S_1^{SM}
\sim A/l_P^2$, where $A$ is the surface area of the black hole. In
other words
for the 'natural' choice of the cut-off parameter $l\sim l_P$ the
one-loop
statistical-mechanical
entropy $S_1^{SM}$ of a black hole is of the same order of magnitude
as the
tree-level Bekenstein-Hawking entropy $S^{BH}$.

We show now that the additional term $\Delta S_1$ in Eq.(\ref{15})
always
exactly compensates the divergence of $S^{SM}_1$ at
$\varepsilon\rightarrow 0$,
so that $S^{TD}_1$ remains finite in this limit.
The key point of the proof is the above mentioned  property of the
renormalized
free energy. Namely,  for $\beta=\beta_{H}=4\pi r_+$  the point
$r=r_+$ is a
regular point of the regular Euclidean manifold with metric
(\ref{2.0b}) and
hence the renormalized partition function $Z_1$  calculated for any
finite
region of this manifold is finite. (We should recall that the black
hole is
surrounded by a boundary, so that the integration must be limited by
$r\le
r_B$.) It implies that $F_1$ calculated for $\beta=\beta_H$ (i.e.,
$F_1=r_+^{-1}{\cal F} (1,0)$) is also finite.
The one-loop contribution $S^{TD}_1$ of a field to the
thermodynamical entropy
of a black hole can be obtained by differentiation of $F_1$ with
respect to the
inverse temperature, provided one  substitutes
$\beta_H=\beta$  into $F_1$ {\em  before} its differentiation. By
using
Eq.(\ref{15}) one gets
\begin{equation}\label{23}
S_1^{TD} =4\pi\beta^2_H{\partial \over \partial \beta_H}\left(
\frac{{\cal
F}(1,\varepsilon)}{\beta_H}\right) _{\varepsilon=0} =
4\pi\left[ -{\cal F}(1,0)+\left({\partial {{\cal F}(1,\varepsilon
)}\over \partial \varepsilon}{\partial {\varepsilon}\over \partial
{\ln
\beta_H}}\right)_{\varepsilon=0} \right] .
\end{equation}
Because for $\beta=\beta_H$ the  free energy $F_1$ does not contain
divergence
at the Euclidean horizon, the quantity in the square brackets of
Eq.(\ref{23})
is finite. It means that  the additional contribution $\Delta S_1$
exactly
compensates the divergent terms of $S_1^{SM}$, so that  the
contribution
$S_1^{TD}$ of the quantum field $\varphi$ to the
thermodynamical entropy of a black hole  is of order of
$O(\varepsilon^0)$.  In particular it means that $S_1^{TD}$  is
independent of
the nature of the cut-off $\varepsilon$, which is assumed in
$S^{SM}_1$ and
which for its calculation requires knowledge of physics at the
Planckian scale.
In other words, the thermodynamical entropy of a black hole is
completely
determined by low energy physics.  $S_1^{TD}$ contains the part which
depends
on $r_B$. This part describes the entropy of thermal gas of quanta of
$\varphi$
field, located outside a black hole within the cavity of size $r_B$.
In
addition $S_1^{TD}$ also contains part independent of $r_B$
describing quantum
corrections to the black hole entropy. For black holes of mass much
larger than
the Planckian mass these corrections are much smaller than $A/l_P^2$
and can be
neglected.
As the result of the above described compensation mechanism the
dynamical
degrees of freedom of
the black hole practically do not contribute to its thermodynamical
entropy
$S^{TD}$, and the latter is defined by the tree-level quantity
$S^{BH}$.

To make the basic idea clearer   we restricted ourselves in the above
discussion by considering a non-rotating black hole. The analysis is
easily
applied to the case of a charged rotating black hole as well as to
their
non-Einsteinian and
n-dimensional generalizations. It is interesting that for black holes
in the
generalized gravitational theories the thermodynamical   and
statistical-mechanical
 entropy may have different dependence on the mass $M$ of a black
hole. In
particular\cite{Frol:92} for a two dimensional dilaton black hole
$S^{BH}=4\pi
M/\sqrt{\lambda}$,  while $S_1^{SM} \sim \ln \varepsilon$.

To summarize it has been shown that the Bekenstein-Hawking entropy
does not
coincide with the statistical-mechanical entropy
$S^{SM}_1=-\mbox{Tr}(\hat{\rho} \ln \hat{\rho} ) $ of a black hole.
The latter
entropy is determined by internal degrees of freedom of the black
hole,
describing  different states which may exist  inside a black hole for
the same
value of its external
parameters. The discrepancy arises  because  in the state of thermal
equilibrium the parameters of  internal degrees of freedom of a black
hole
depend on the
temperature of the system in the universal way. This results in the
universal
cancellation of all those contributions to the thermodynamical
entropy which
depend on the particular properties and number of  fields. That is
why the
thermodynamical entropy of  black holes in  Einstein's  theory is
always
$S^{BH}$.

\section{Acknowledgements}

This work was supported by the Natural Sciences and Engineering
Research
Council of Canada.


\begin{thebibliography}{00}

\bibitem{Beke:72} J.~D.Bekenstein, Nuov.Cim.Lett. {\bf 4} (1972) 737.
\bibitem{Beke:73} J.~D.~Bekenstein,  Phys.Rev. {\bf D7} (1973)  2333.
\bibitem{BaCaHa:73} J.~M.~Bardeen, B.~Carter, and S.~W.~Hawking,
Comm.Math.Phys.{\bf 31} (1973) 181.
\bibitem{Beke:74} J.~D.~Bekenstein,  Phys.Rev. {\bf D9}  (1974)
3292.
\bibitem{ThPrMa:86} K.~S.~Thorne, W.~H.~Zurek, and R.~H.~Price, \  in
{\em
Black Holes: The Membrane Paradigm }, edited by K.~S.~Thorne,
R.~H.~Price, and
D.~A.~MacDonald (Yale University Press, New Haven, 1986), p.280.
\bibitem{NoFr:89} I.~Novikov and V.~Frolov, {\em Physics of Black
Holes.}
(Kluwer Academic Publ., Dordrech-Boston-London), 1989.
\bibitem{Wald:92} R.~W.~Wald, In:{\em Black Hole Physics}
(Eds.V.~DeSabbata and
Z.~Zhang), (Kluwer Academic Publ., Dordrech-Boston-London), 1992).
\bibitem{FrPa:93} V.~Frolov and D.~N.~Page, Phys.Rev.Lett. {\bf 71}
(1993)
3902.
\bibitem{ZuTh:85}  W.~H.~Zurek and K.~S.~Thorne, Phys.Rev.Lett. {\bf
54} (1985)
2171.
\bibitem{GiHa:76} G.~W.~Gibbons and S.~W.~Hawking,  Phys.Rev. {\bf
D15}
(1976) 2752.
\bibitem{Hawk:79} S.~W.~Hawking, In: {\em General Relativity: An
Einstein
Centenary Survey.} (eds. S.~W.~Hawking and W.~Israel), Cambridge
Univ.Press,
Cambridge, 1979.
\bibitem{York:86} J.~W.~York,  Phys.Rev. {\bf D33}  (1986) 2092.
\bibitem{BrBrWhYo:90} H.~W.~Braden, J.~D.~Brown, B.~F.~Whiting, and
J.~W.~Jork,
Phys.Rev. {\bf D42}  (1990) 3376.
\bibitem{Alle:86} B.~Allen, Phys.Rev. {\bf D33} (1986) 3640.
\bibitem{Hoof:85} G.~'t Hooft, Nucl.Phys. {\bf B256} (1985) 727.
\bibitem{FrNo:93} V.~Frolov and I.~Novikov, Phys.Rev. {\bf D48}
(1993) 4545.
\bibitem{CaTe:93} S.~Carlip and C.~Teitelboim, Preprint gr-qc/9312002
(1993).
\bibitem{SuUg:94} L.~Susskind and J.~Uglum, Phys.Rev. {\bf D50}
(1994) 2700.
\bibitem{GaGiSt:94} D.~Garfinkle, S.~B.~Giddings, and A.~Strominger,
Phys.Rev.
{\bf D49} (1994) 958.
\bibitem{BaFrZe:94}  A.~I.~Barvinsky, V.~P.~Frolov, and
A.~I.~Zelnikov,
preprint  Thy 13-94, gr-qc/9404036 (1994), (to appear in Phys.Rev.D).
\bibitem{Beke:94}   J.~D.~Bekenstein, preprint   gr-qc/9409015
(1994).
\bibitem{Hawk:75} S.~W.~Hawking, Comm.Math.Phys.
{\bf 43},   (1975) 199.
\bibitem{Frol:94}   V.~Frolov, preprint  Thy 22-94, gr-qc/9406037
(1994).
\bibitem{EvHaUmYa:93} T.~S.~Evans, I.~Hardman, H.~Umezawa and
Y.~Yamanaka,
Fortschr.Phys. {\bf 41}  (1993)151.
\bibitem{BrYo:93} J.~D.~Brown and J.~W.~Jork, Phys.Rev. {\bf D47}
(1993)1420.
\bibitem{DoKe:78} J.~S.~Dowker and G.~Kennedy, J.Phys. {\bf A 11}
(1978) 895.
\bibitem{Frol:92} V.~P.~Frolov,  Phys.Rev. {\bf D46} (1992) 5383.
\end{thebibliography}
\end{document}